\documentclass{article}\usepackage[]{graphicx}\usepackage[]{xcolor}
\makeatletter
\def\maxwidth{ %
  \ifdim\Gin@nat@width>\linewidth
    \linewidth
  \else
    \Gin@nat@width
  \fi
}
\makeatother

\definecolor{fgcolor}{rgb}{0.345, 0.345, 0.345}

\usepackage{framed}
\makeatletter
 {\par\unskip\endMakeFramed%
 \at@end@of@kframe}
\makeatother

\definecolor{shadecolor}{rgb}{.97, .97, .97}
\definecolor{messagecolor}{rgb}{0, 0, 0}
\definecolor{warningcolor}{rgb}{1, 0, 1}
\definecolor{errorcolor}{rgb}{1, 0, 0}
\newenvironment{knitrout}{}{} % an empty environment to be redefined in TeX

\usepackage{alltt}
\input{mytheme.sty}
\usepackage{orcidlink}
\usepackage{amsmath}
\usepackage{tikz}
\usepackage{booktabs}
\usepackage[sc]{mathpazo}
\newcommand{\pttr}{p_{\tiny \mbox{2TR}}}
\newcommand{\TOE}{Type-I error }

\usepackage[most]{tcolorbox}
\usepackage{multirow}

\newtcbtheorem{Summary}{\bfseries Summary}{enhanced,drop shadow={black!50!white},
  coltitle=black,
  top=0.3in,
  attach boxed title to top left=
  {xshift=1.5em,yshift=-\tcboxedtitleheight/2},
  boxed title style={size=small,colback=pink}
}{summary}

\newtcolorbox[auto counter]{summary}[1][]{title={\bfseries Box~\thetcbcounter},enhanced,drop shadow={black!50!white},
  coltitle=black,
  top=0.3in,
  attach boxed title to top left=
  {xshift=1.5em,yshift=-\tcboxedtitleheight/2},
  boxed title style={size=small,colback=pink},#1}

\title{\bf{The assessment of replicability using the sum of \textit{p}-values}}
\author{Leonhard Held \orcidlink{0000-0002-8686-5325}, Samuel Pawel \orcidlink{0000-0003-2779-320X}, and Charlotte Micheloud \orcidlink{0000-0002-4995-4505}\\
  Epidemiology, Biostatistics
  and Prevention Institute (EBPI)\\ 
  and Center for Reproducible Science (CRS) \\
  University of Zurich\\ Hirschengraben 84,
  8001 Zurich, Switzerland\\ \small Email: \texttt{\{leonhard.held,samuel.pawel,charlotte.micheloud\}@uzh.ch}}
\IfFileExists{upquote.sty}{\usepackage{upquote}}{}
\begin{document}

\maketitle

\vspace{-.5cm}
\begin{center}
\begin{minipage}{12cm}
\paragraph{Abstract} 
Statistical significance of both the original and the replication
study is a commonly used criterion to assess replication attempts,
also known as the two-trials rule in drug development.  However,
replication studies are sometimes conducted although the original
study is non-significant, in which case Type-I error rate control
across both studies is no longer guaranteed.  We propose an
alternative method to assess replicability using the sum of
$p$-values from the two studies.  The approach provides a combined
$p$-value and can be calibrated to control the overall Type-I error
rate at the same level as the two-trials rule but allows for
replication success even if the original study is non-significant.
The unweighted version requires a less restrictive level of
significance at replication if the original study is already
convincing which facilitates sample size reductions of up to 10\%.
Downweighting the original study accounts for possible bias and
requires a more stringent significance level and larger samples sizes
at replication. Data from four large-scale replication
projects are used to illustrate
and compare the proposed method with the two-trials rule,
meta-analysis and Fisher's combination method.
\\[0.5cm]
 
\end{minipage}
\end{center}

\newpage
\section{Introduction}\label{sec:intro}

Replication studies are increasingly conducted in various fields to
assess the replicability of original findings. %% This has led researchers to
%% propose novel ways to design and analyse such studies \citep[among
%%   others]{Simonsohn2015, Anderson2017, Hedges2019b, Bonett2020,
%%   Held2020}.
While `replicability' is intuitively understood as the ability to obtain
consistent results when a study is repeated with new subjects \citep{NSF2019},
there is no agreed-upon quantitative definition of replicability or `replication
success'. In practice, a variety of different approaches are used that capture
different intuitions about what it means for a replication to be successful. For
example, metrics such as the $Q$-test or prediction intervals quantify the
statistical compatibility of the effect estimates from original and replication
studies \citep{Patil2016, Hedges2019b} while a meta-analysis of original and
replication results provides an assessment of the evidence pooled across both
studies \citep{Muradchanian2023}.

Perhaps the most commonly used success criterion is statistical significance of the
replication study with an effect estimate in the same direction as in the
original study \citep{open2015,Camerer2016,Camerer2018,Cova2018,Errington2021}.
Positive original findings are usually also based on significance, and the
%% If,
%% additionally, significance of the original study is required to
%% justify the conduct a replication study in the first place, this
procedure is then analogous to the `two-trials rule' in drug development
\citep[Sec.~12.2.8]{Senn2021}. The intuition behind this approach is that each
study individually should provide sufficient evidence for a claim, which is not
guaranteed with other approaches such as meta-analysis or the $Q$-test.
In practice, the standard two-sided significance level 0.05 is often
used and so replication success with the two-trials rule is achieved
if both one-sided $p$-values $p_o$ and $p_r$ from the original and
replication study, respectively, are smaller than $\alpha = 0.05/2 = 
0.025$. The one-sided formulation is convenient to
ensure that the effect estimates from the two studies are in the same direction. 

The success condition of the two-trials rule $\max\{p_o, p_r\} \leq \alpha$ can be rewritten as 
  \begin{eqnarray}\label{eq:p2tr}
  \pttr = \max\{p_o, p_r\}^2 \leq \alpha^2\, ,
  \end{eqnarray}
where the combined $p$-value $\pttr$ of the two-trials rule turns out to be
  {\em valid} \citep{Greenland2019}, \ie uniformly distributed under the intersection null
  hypothesis that both the true effect $\theta_o$ in the original
  study and the true effect $\theta_r$ in the replication study are null. The two-trials rule therefore controls the
\emph{overall Type-I error rate} at level $\alpha^2$.  For
  $\alpha=0.025$, the probability to incorrectly declare replication
  success under the intersection null hypothesis is 
  $0.025^2=0.000625$.
%
%% For sample size
%% planning, the replication sample size is usually chosen to detect the
%% original effect estimate at significance level 0.025 with sufficiently
%% large power, say 80\% or 90\%.

Despite its simplicity, in the replication setting the two-trials rule
has important limitations.  First of all, the 'double dichotomization'
at $\alpha = 0.025$ can lead to conclusions which seem
counterintuitive. For example, the two-trials rule is not fulfilled if
$p_o = 0.026$ and $p_r = 0.001$ (or {\em vice versa}), but
it is fulfilled when both $p_o$ and $p_r$ are 0.024, although
there is less evidence for an effect in the second case \citep{BenjaminBerger2019}.
The first example also illustrates that replication
success is impossible if the original study just missed statistical
significance, no matter how convincing the replication study is.
However, some replication projects do interpret non-significant
original studies as positive findings and try to replicate them. In
the \emph{Reproducibility Project: Psychology}
\citep{open2015}, four 'positive', but non-significant original effects
have been included, all with (one-sided) $p$-values between 0.025 and
0.03. In the \emph{Experimental Economics Replication Project} \citep{Camerer2016}, two 'positive', but non-significant effects have been
included, one with $p_o=0.027$ \citep{AmbrusGreiner2012}, one with an even larger $p$-value
$p_o=0.035$ \citep{Kuziemko_etal2014}.
%% \todo[inline]{SP: Looking at the table in the appendix from Camerer (2016), I see that the original two-sided p-value from Ambrus and Greiner is 0.057, so slightly different from the one based on normal approximation that you report. The one from Kuziemko et al. is the same, though.}
%However,
Strict application of the two-trials rule would inflate the overall Type-I error
rate and reduce trust in the replication finding.

The past and ongoing debate on the use and misuse of $p$-values \citep{Wasserstein2016,Colquhoun:2017,Wasserstein_etal2019}
has led various researchers to advocate for a
quantitative interpretation of $p$-values as measures of the strength
of evidence \citep{HeldOtt2018} or divergence
\citep{Greenland2023}. This implies that an original study with a relatively
large $p$-value %%(close to the traditional two-sided $0.05$ significance
%% threshold)
carries only suggestive evidence against the null hypothesis of no
effect \citep{Benjamin_etal2018} and requires a more convincing replication
result for confirmation than an original study with a smaller $p$-value. There
is in fact empirical evidence that studies with small $p$-values tend to
replicate better than studies with only suggestive evidence. For example,
several large-scale replication projects found strong negative Spearman
correlations between original $p$-values and the corresponding (two-sided) replication
$p$-values being smaller than 0.05 \citep{OSC2015, Camerer2016, Camerer2018}.
% \citep[]{Held2019,PawelHeld2020}.\todo[inline]{SP: isn't this also investigated in the replication projects themselves?}
\citet{Held2019} has shown that more stringent significance levels %% \citep[extending the][proposal]{Benjamin_etal2018} 
have improved performance in predicting replication success in an application to the data from the  \citet{OSC2015} project.  
The two-trials rule, however, requires the same level of evidence
at replication no matter what level of evidence the claim of
the original discovery had, as long as it was significant. 

A promising alternative is to summarize the total evidence from the
two $p$-values with a combined $p$-value, but different from the one based on the two-trials rule \eqref{eq:p2tr}. A large number of valid $p$-value
combination methods are available
\citep{HedgesOlkin1985,cousins2007annotated} and the two-trials rule
can serve as a benchmark as it provides the success
threshold $\alpha^2$ to ensure appropriate control of the overall Type-I
error rate \citep{Fisher1999, Shun_etal2005,Rosenkranz2023}. %% Many of them ensuring 
%% that the larger the original $p$-values was, the more strict we have to be at 
%% replication to flag
%% replication success. 
Perhaps most prominent  is Fisher's combination method \citep{Fisher1935}, where replication success at level 
$\alpha^2$
is achieved if the product of the $p$-values $p_o p_r \leq c_F = \exp\{-0.5\chi^2_4(1 - \alpha^2)\}$, 
where $\chi^2_\nu(\cdot)$ denotes the quantile
function of the $\chi^2$-distribution with $\nu$ degrees of freedom. 
The corresponding combined $p$-value is
\begin{eqnarray}
p_F = 1 - \Pr\left(X \leq -2 \log \{p_o p_r\}\right) \, ,
\end{eqnarray}
where $X$ follows a $\chi^2$ distribution with 4 degrees of freedom, so 
replication success would be declared if $p_F \leq \alpha^2$. However,
the goal of replication studies is to confirm an original (positive)
finding, but replication success with Fisher's method can be achieved
even if one of the $p$-values is very large.  In fact, replication
  success at level $\alpha^2=0.025^2$ is guaranteed if $p_o < c_F \approx
  0.00006$, no matter
  how large $p_r$ is, so Fisher's method may not even require a
  replication study to confirm a claim of a new discovery, which is clearly undesirable. 
%%  This method
%%  is therefore clearly inappropriate as a criterion for replication success. % alternative to the two-trials rule?

An alternative approach used often in replication projects
  \citep{open2015,Camerer2016,Camerer2018,Errington2021} is to calculate a 
combined $p$-value $p_{\mbox{\tiny MA}}$ with a fixed-effect
meta-analysis, %% of original and replication effect estimates, 
also known as the `pooled-trials rule' in drug development \citep[Section 12.2.8]{Senn2021}.
Let $\Phi(\cdot)$ denote the standard normal cumulative distribution function 
and $\Phi^{-1}(\cdot)$ the corresponding quantile function. 
The meta-analytic $p$-value $p_{\mbox{\tiny MA}} = 1 - \Phi(z_{\mbox{\tiny MA}})$
is identical to the combined 
$p$-value from weighted Stouffer's method \citep{Stouffer1949,cousins2007annotated}, where
\begin{equation}
  %%p_{\mbox{\tiny MA}}\, ,
%  \mbox{ where } 
z_{\mbox{\tiny MA}} = \frac{\sigma_r \, z_o + \sigma_o \, z_r}{\sqrt{\sigma_o^2 + \sigma_r^2}} \,  \label{eq:pMA}
\end{equation}
is the weighted average of the original and 
replication $z$-values $z_o = \Phi^{-1}(1 - p_o)$ and $z_r = \Phi^{-1}(1 - p_r)$
with weights proportional to the standard errors $\sigma_r$ and $\sigma_o$ of
$\hat \theta_r$ and $\hat \theta_o$, respectively.
%% 1 and the variance ratio $c = \sigma_o^2/\sigma_r^2$ (where
%% $\sigma^2_o$ and $\sigma_r^2$ denotes the squared standard error of
%% $\hat \theta_o$ and $\hat \theta_r$, respectively). 
%
%%Again %% To ensure the same overall T1E rate as the two-trials rule, 
Again, the criterion $p_{\mbox{\tiny MA}} \leq \alpha^2$ has been
used to assess replication success
\citep{Fisher1999,Maca2002,Shun_etal2005,Freuli2023}, although a threshold
larger than 0.000625 (\eg two-sided 0.005, or even 0.05) is often used
in applications.  But it is immediate from~\eqref{eq:pMA} that
meta-analysis has similar problems as Fisher's method as
$z_{\mbox{\tiny MA}}$ can become large (and hence $p_{\mbox{\tiny MA}}$
small) even if one of the two underlying $z$-values $z_o$ and $z_r$ is
small or even negative (and hence the corresponding $p$-value
large). Recently \citet{Muradchanian2023} have conducted a study to
investigate meta-analysis as a replication success metric. They also
conclude that meta-analysis is an inappropriate tool if one wants to
evaluate whether the replication result is in line with the original
result.
%% ``quantifying replication success using meta-analysis resulted in the conclusion
%% where the replication was deemed a success regardless of the results of the replication study''.

In what follows we propose a particular $p$-value combination method based on the 
  sum of $p$-values which requires both studies to be convincing, similar to the two-trials rule:
Edgington's method
\citep{Edgington1972}, recently rediscovered by \citet{Held2024} in the 
context of drug regulation. %%The
%%method can be calibrated to control the overall T1E rate at level
%%$\alpha^2$. However, i
For overall Type-I error control at level $\alpha^2=0.025^2$, success is achieved if the sum of $p$-values is
smaller than $\sqrt{2} \cdot 0.025 \approx 0.035$, so possible if
either $p_o$ or $p_r$ is not significant, as long as they are both
smaller than $0.035$. As such, the method can be used in the scenarios encountered in previous replication projects where an original study was borderline non-significant.  %% Success is possible if either $p_o$ or $p_r$ is
%% not significant, as long as they are both smaller than $0.035$. 
%% Sample
%% size planning of the replication study is hence also possible for non-significant original
%% studies.
We also derive a weighted version of Edgington's method that
treats original and replication differently.  Giving twice the weight
to the replication study results in the necessary (but not
sufficient) success conditions $p_o \leq 2 \, \alpha$ and $p_r \leq
\alpha$. Both version of Edgington's method reach opposite conclusions
than the two-trials rule for the two examples introduced earlier:
replication success at level $0.025^2$ is declared if $p_o = 0.026$ and $p_r =
0.001$, but not if $p_o$ and $p_r$ are 0.024. The approach is summarized in
Box \ref{box.edgington}.
%% \todo[inline]{SP: Perhaps one could also say that both version control the T1E at the same level but reach opposite conclusions as the two-trials rule?}
%%\todo[inline]{SP: would also be interesting to know the p-values from Fisher and Stouffer here}
%

%\subsection{Related methodology in adaptive clinical trials}

\begin{summary}{}
  \textbf{Assessing Replicability Using the Sum of \textit{P}-Values} \\
  Overall Type-I error control at level $\alpha^2=0.025^2$\\
  Input: One-sided $p$-values $p_o$ and $p_r$ from original and replication study %% with $p_o + w_r/w_o p_r \leq 1$
\begin{center}
  \begin{tabular}{lcc}
   & Unweighted & Weighted \\ \hline
  Weights & $w_o=1$, $w_r=1$ & $w_o=1$, $w_r=2$ \\ \hline
  Replication success &
  \begin{multirow}{2}{*}%%{test}
      {$\begin{array}{rcl}
      p_o + p_r & \leq & \sqrt{2} \, \alpha \\
      & \approx & 0.035 \\
      \end{array}$}
\end{multirow}
  &
    \begin{multirow}{2}{*}%%{test}
      {$\begin{array}{rcl}
      p_o + 2\, p_r & \leq & {2} \, \alpha \\
      & = & 0.05 \\
      \end{array}$}
\end{multirow} \\
%% $p_o + 2 \, p_r \leq {2} \, \alpha = 0.05$\\ 
    criterion &  \\ \hline
    Significance level for &  \begin{multirow}{2}{*}%%{test}
      {$\begin{array}{rl}
      & \sqrt{2} \, \alpha - p_o   \\
      \approx & 0.035 - p_o\\
      \end{array}$}
\end{multirow} &  \begin{multirow}{2}{*}%%{test}
      {$\begin{array}{rl}
      & \alpha - p_o/2   \\
      = & 0.025 - p_o/2 \\
      \end{array}$}
\end{multirow} \\
  replication study &  & \\ \hline
    Combined $p$-value & $p_E = (p_o + p_r)^2/2$ & $p_{E_w} = (p_o + 2 \, p_r)^2/4$ \\ 
& (for $p_o + p_r \leq 1$) & (for $p_o + 2\, p_r \leq 1$) \\ \hline
  \end{tabular}
  \end{center}
%%   \begin{itemize}
%%   \item Parameters:
%%     \begin{itemize}
%%     \item Unweighted: Weights $w_o=1$ and $w_r=1$ for original and replication study
%%     \item Weighted: Weights $w_o=1$ and $w_r=2$ for original and replication study
%% \end{itemize}
%%     \item   Analysis: 
%%   \begin{itemize}
%%   \item Unweighted: Flag success if $p_o + p_r \leq 0.035$
%%   \item Weighted: Flag success if $p_o + 2 \, p_r \leq 0.05$
%%   \end{itemize}
%%   \item   Sample size planning of replication study: 
%%   \begin{itemize}
%%   \item Unweighted: Use significance level $0.035 - p_o$
%%   \item Weighted: Use significance level $0.025 - p_o/2$
%%   \end{itemize}
%% \item[]
  Other choices can be made for $\alpha$ and the weights $w_o$ and $w_r$
%% \end{itemize}
  \label{box.edgington}
\end{summary}

The rest of the paper is organized as follows. 
Edgington's method is described in Section~\ref{sec:Edg} and extended
to include weights in Section~\ref{sec:EdgW}. 
In Section~\ref{sec:pT1E}, we
discuss and compare the \emph{conditional Type-I error rate} and the
\emph{project power} of the two-trials rule, meta-analysis, and
Fisher's and Edgington's methods.
%% \todo[inline]{SP: Type-I error was already mentioned before several times, should the acronym be introduced at the first occurrence?}
The conditional \TOE rate is the
probability, given the original study result, to incorrectly declare
replication success when the true effect is null at replication, while
the project power is the probability to correctly declare replication
success over both studies in combination.  The different methods are
applied to the data from four large-scale replication projects
in~Section~\ref{sec:appli}, and sample size planning of the
replication study is described in Section~\ref{sec:design}.
Section \ref{sec:moreThanOne} describes extensions to more than one replication study. 
Finally,
some discussion is provided in Section~\ref{sec:discussion}.

\section{Additive combination of \textit{p}-values}

\subsection{Edgington's method}\label{sec:Edg}
\citet{Edgington1972} developed a method to combine $p$-values 
by adding them. %%, recently 
%%rediscovered as an alternative to the two-trials rule for
%%drug approval \citep{Held2024}.
Here we investigate the use of Edgington's method 
in the replication setting, where results from one original and one replication 
study are available.
Under the intersection null hypothesis, the sum of the two $p$-values
\begin{eqnarray}\label{eq:E}
E =  p_o + p_r
\end{eqnarray}
follows the Irwin-Hall distribution with parameter $n=2$  \citep{Irwin1927, Hall1927}, 
\ie $E \sim$ IH(2). The cumulative distribution function of the 
Irwin-Hall distribution can then be used to  calculate a valid combined 
$p$-value 
\begin{eqnarray}\label{eq:pE}
  p_E = \Pr(\mbox{IH}(2) \leq E) = \left\{ \begin{array}{rl} E^2/2 &
    \mbox{ if $0 < E\leq 1$} \, , \\
    -1 + 2\,E - E^2/2 & \mbox{ if $1 < E \leq 2$} \, . \end{array} \right.
\end{eqnarray}
The succes condition $p_E \leq \alpha^2$ can be expressed in terms of the 
the $p$-value sum $E \leq b$, where
the critical value is $b = \sqrt{2} \, \alpha$. At the standard $\alpha = 0.025$, 
the critical value is hence $b \approx 0.035$.
The threshold $b$ can be considered 
as the available budget for the two $p$-values $p_o$ and $p_r$, and
this implies that replication success is possible for a non-significant 
original (or replication) $p$-value as long as $p_o$ (or $p_r$) 
$ < b = 0.035$. 
The sum of the $p$-values for the two
examples presented in the introduction is $E = 0.026 + 0.001 = 0.027$ 
and $E = 0.024 + 0.024 = 0.048$, respectively, so replication success is declared 
in the first example but not the second, in contrast 
to the two-trials rule but in accordance to intuition. The corresponding combination $p$-values \eqref{eq:pE}
are $p_E = 0.0004$ and $p_E = 0.001$, respectively, the first smaller and the second larger than the threshold $\alpha^2 = 0.000625$. 

\subsection{Weighted version}\label{sec:EdgW}

An interesting extension of Edgington's method is to include weights, for example
%% as in the weighted Stouffer's method. One option would be to use
%% weights 
proportional to the precision of the studies, as in weighted
Stouffer's method. However, in the replication setting one might
want to downweight the original and upweight the replication study,
for example with weights $1/3$ and $2/3$, respectively. This would
address concerns that original studies may be subject to questionable
research practice and hence prone to bias. Consider the weighted sum
of $p$-values
\begin{eqnarray}\label{eq:Ew}
E_w =  w_o p_o + w_r p_r
\end{eqnarray}
with positive weights $w_o \leq w_r$, then we show in Appendix
\ref{app:weightedSum} that the corresponding combination $p$-value is
\begin{eqnarray}\label{eq:pEw}
p_{E_w} =  \left\{ \begin{array}{rl} \frac{E_w^2}{2\, w_o w_r} &    \mbox{ if $0 < E_w \leq w_o$} \, , \\
    \frac{1}{w_r} \left(E_w - \frac{w_o}{2} \right)& \mbox{ if $w_o < E_w \leq w_r$} \, ,  \\
     1 + \frac{1}{w_o w_r}\left(E_w(w_o + w_r) - \frac{(w_o + w_r)^{2}}{2} - \frac{E_w^{2}}{2} \right)
    %% - 1 - \frac{1}{2\, w_o w_r}\left((w_o-w_r)^2 - 2(w_o + w_r)E_w + {E_w^{2}} \right)
    & \mbox{ if $w_r < E_w \leq w_o + w_r$} \, .
  \end{array} \right.
\end{eqnarray}
Note that \eqref{eq:pEw} reduces to \eqref{eq:pE} for $w_o=w_r=1$, and is invariant under multiplication of the weights $(w_o, w_r)$ % \rightarrow (v \, w_o, v \, w_r)$
with a positive constant. The $p$-value  \eqref{eq:pEw} therefore only depends on the weight ratio $\tilde w = w_r/w_o$ of the 
replication to the original weight.
Setting the first line of \eqref{eq:pEw} equal to $\alpha^2$ gives the available budget $b_w =
\sqrt{2 w_o w_r} \, \alpha$  for $E_w \leq w_o$.

In the following we will use the weights $w_o=1$ and $w_r=2$,
although other choices can be made, of course. 
Then 
%% \begin{eqnarray}\label{eq:Ew21}
%% E_w =  w_o p_o + w_r p_r
%% \end{eqnarray}
%% with positive weights $w_o \leq w_r$, then we show in Appendix
%% \ref{app:weightedSum} that the corresponding combination $p$-value is
\begin{eqnarray}\label{eq:pEw12}
p_{E_w} =  \left\{ \begin{array}{rl} {E_w^2}/{4} &    \mbox{ if $0 < E_w \leq 1$} \, , \\
    E_w/2 - {1}/{4} & \mbox{ if $1 < E_w \leq 2$} \, ,  \\
    - \frac{5}{4} + \frac{3}{2} E_w - {E_w^{2}}/{4} 
%%    + \frac{1}{2}\left(3 \, E_w - \frac{9}{2} - \frac{E_w^{2}}{2} \right)
    %% - 1 - \frac{1}{2\, w_o w_r}\left((w_o-w_r)^2 - 2(w_o + w_r)E_w + {E_w^{2}} \right)
    & \mbox{ if $2 < E_w \leq 3$} \, .
  \end{array} \right.
\end{eqnarray}
%%\todo[inline]{SP: Maybe the formatting of fractions could be the same everywhere? i.e., use $1/4$ or $\frac{1}{4}$ but not a combination?}
For $E_w \leq 1$ with small enough $p_o < p_r$, we have $p_{E_w} = (p_o +
2 \, p_r)^2/4 \approx p_r^2 = p_{2TR}$ from \eqref{eq:p2tr}, so weighted
Edgington will behave similar to the two-trials rule, whereas the $p$-value from the
unweighted version will then be roughly half as large as  $p_{2TR}$:  $p_{E} = (p_o +
p_r)^2/2 \approx p_r^2/2 = p_{2TR}/2$.%%  Differences between $p_{E_w}$ and $p_{2TR}$ will be more pronounced
%% for larger $p_o$. 

The success condition $p_{E_w} = E_w^2/4 \leq \alpha^2$ can be re-written
as $E_w = p_o + 2\, p_r \leq 2 \, \alpha$, so doubling the
replication weight increases the budget from $\sqrt{2} \, \alpha$ to $2
\, \alpha$ and for $\alpha=0.025$ it will be possible to successfully
replicate original studies with $p_o \leq 0.05$.  However, the
$p$-value of the replication study now counts twice in $E_w$, so
replication success is impossible if $p_r > \alpha$, just as with the
two-trials rule.  For example, the original study by
\citet{Kuziemko_etal2014} mentioned in Section \ref{sec:intro} had a
quite large $p$-value: 
$p_o=0.035$. Conducting a replication would be pointless if analysis
is based on the two-trials rule or even unweighted Edgington. However,
replication success would be still possible with weighted Edgington,
but the replication study has to be quite convincing to achieve
success ($p_r < 0.025 - 0.035/2 = 0.0075$).

\section{Operating characteristics}\label{sec:pT1E}
As discussed before, all methods considered control the overall Type-I error rate at level $\alpha^2$. 
We will now look at two other operating characteristics: conditional Type-I error rate and project power. 

\subsection{Conditional Type-I error rate}\label{sec:CT1E}

The success condition on the replication $p$-value $p_r$ with the two-trials rule 
is always $p_r \leq \alpha$, regardless of the original study result (as long as $p_o \leq \alpha$ holds). In contrast, 
with Edgington's and Fisher's methods as well as the meta-analysis criterion
the required value of $p_r$ depends on the 
original $p$-value $p_o$. This condition is 
\begin{eqnarray}\label{eq:condE}
p_r \leq b - p_o \, 
\end{eqnarray}
for Edgington's method,
\begin{eqnarray}\label{eq:condEw}
p_r \leq (b_w - w_o p_o)/w_r \, 
\end{eqnarray}
for weighted Edgington's method,
\begin{eqnarray}\label{eq:condF}
p_r \leq \min\{c_F/p_o, 1\} \, 
\end{eqnarray}
for Fisher's method, and
  \begin{eqnarray}\label{eq:condS}
  p_r \leq 1 - \Phi\left\{\frac{\Phi^{-1}(1 - \alpha^2)\sqrt{c + 1} - z_o}{\sqrt{c}}\right\} \, 
  \end{eqnarray}
for the meta-analysis criterion, which depends on the variance ratio $c = \sigma_o^2/\sigma_r^2$.
For a very convincing original study (where $p_o \to 0$, equivalently $z_o \to \infty$), the right hand-side
in~\eqref{eq:condE} tends to $b = \sqrt{2}\, \alpha$, in ~\eqref{eq:condEw}
to $b_w/w_r = \sqrt{2 / \tilde w} \, \alpha$, while the right-hand side in
in~\eqref{eq:condF} and~\eqref{eq:condS} converges to $1$. 
Replication success can thus be achieved with Fisher's method and 
the meta-analysis criterion even if the replication 
$p$-value is very large, while this cannot happen with the two-trials rule 
and Edgington's method.

Of particular interest in the replication setting is the conditional
  \TOE rate, the probability, given the original study result,  that the replication study flags success
  although there is no true effect at replication ($\theta_r=0$). For $\alpha = 0.025$ the conditonal \TOE rate
is $\alpha = 0.025$, 
$b - p_o < b = 0.035$
and $(b_w - w_o p_o)/w_r < b_w/w_r = 0.025$
with the two-trials rule and the unweighted and weighted Edgington's method, respectively, but
can become very large with Fisher's method and the meta-analysis criterion, if $p_o$ is small.
For example, suppose the original study had a $p$-value of $p_o =
0.001$. %% , the conditional Type-I error rate of the
%% two-trials rule.
The conditional
\TOE rate of Edgington's method  at the standard
$\alpha=0.025$ level then is %% 2.5\% and
3.4\% (unweighted) and 2.45\% (weighted), respectively. With Fisher's method and meta-analysis (for $c=1$),
the conditonal \TOE rate is 5.8\% and
7.0\%, respectively.
Now suppose the original study had a $p$-value of $p_o =
0.0001$. The conditional
\TOE rate then is %% 2.5\% and
3.53\%  (unweighted) and 2.495\% (weighted) with %% the two-trials rule and
Edgington's method, so only slightly larger. However, with Fisher's method and meta-analysis,
the conditonal \TOE rate increases drastically to 58.1\% and
19.9\%, respectively.

\begin{figure}[!h]
  \centering
\begin{knitrout}
\definecolor{shadecolor}{rgb}{0.969, 0.969, 0.969}\color{fgcolor}
\includegraphics[width=\maxwidth]{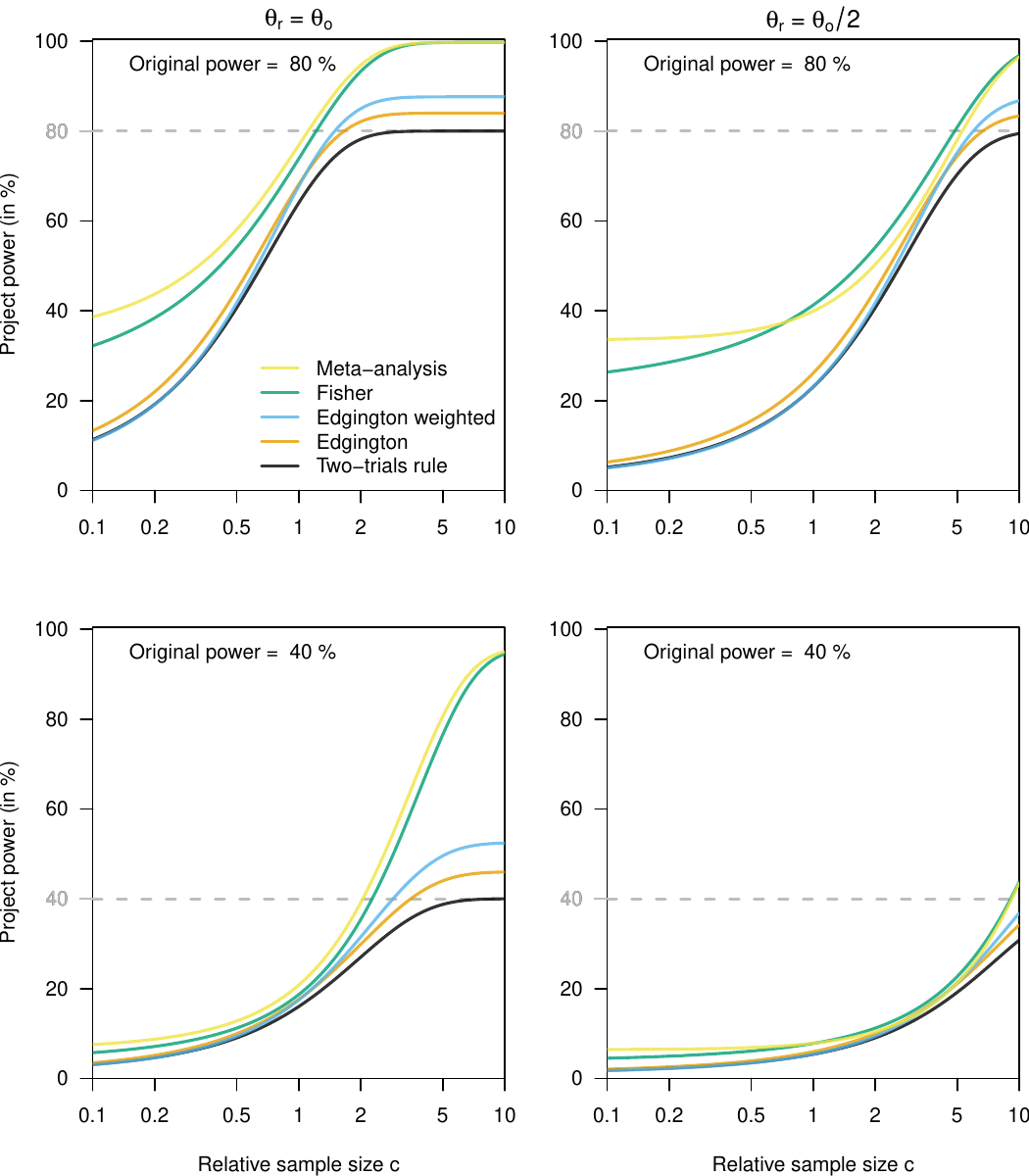} 
\end{knitrout}
\caption{Project power of the two-trials rule, Edgington's (both unweighted and weighted) and Fisher's methods,
and the meta-analysis criterion
  as a function of the relative sample size $c = n_r/n_o$, assuming that 
  the true effect sizes $\theta_o$ and $\theta_r$ are the same (left),
  or that the true replication effect size is half the true original effect size
  $\theta_r = \theta_o/2$ (right).
  The calculations assume that the original study has a power of 80\% (top)
  or 40\% (bottom) to
  detect the assumed true effect size at significance level $\alpha$.}
\label{fig:PP}
\end{figure}

\subsection{Project power}\label{sec:pp}
The project power is the probability to correctly declare replication
success over both studies in combination when both underlying effects
are non-null. Most original studies are designed to have 80\% power
to detect the assumed true effect size at significance level $\alpha$, but
the power can be considerably lower in reality \citep{Turner2013, DumasMallet2017}.
In Figure~\ref{fig:PP} we consider the project power with an original
power of $80$\% (top) and 40\% (bottom),
under the assumption that the true effect sizes are the same ($\theta_r = \theta_o$, left)
or that the true replication effect size is half the true original effect size ($\theta_r =
\theta_o/2$, right). The latter case reflects the shrinkage
of effect estimates often encountered in replication projects.
  
In
contrast to the conditional \TOE rate, the project power of all methods
depends on the original power to detect $\theta_o$ and the
variance ratio $c$, which can often be 
interpreted as the relative sample size $c = n_r/n_o$ (replication to original), 
see Appendix~\ref{app:pp} for the derivations. The project power 
of the two-trials rule cannot become larger than the power in the original 
study to detect the true effect size, %% $80$\% , 
%% respectively $power_pp[2]*100$\%,
see Figure~\ref{fig:PP}.
%% In contrast, t
The project power of Edgington's method 
is either essentially identical (for small $c$ and the weighted version), or otherwise larger than the project power of the two-trials rule with limit
84\% (unweighted) and 87.6\% (weighted), respectively,  for $c \rightarrow \infty$ and an original power of $80$\%. The corresponding values are  
46\% (unweighted) and 52.5\% (weighted)
for an original power of $40$\%.
%% The increase in project power is between round(min(PPdiff),1) and 
%% round(max(PPdiff),1)
%% percentage points in the case where $\theta_r = \theta_o$ 
%% and between round(min(PPdiff_05),1) and 
%% round(max(PPdiff_05),1)
%% percentage points in the case where $\theta_r = \theta_o/f$ 
%% for the range considered 
%% ($min(c_seq) \leq c \leq max(c_seq)$).
The project power of Fisher's method and the meta-analysis criterion is larger than 
the project power of both the two-trials rule and Edgington's method %% for every value 
%% of the relative sample size $c$,
and reaches values close to 100\% in the case $\theta_r = \theta_o$ even 
for a low original power of 40\%. However,  the price to pay
is a considerable increase in conditional \TOE rate, as discussed in Section \ref{sec:CT1E}.
% Results are qualitatively similar with other values of the 
% original power, see Figure~\ref{fig:PP2} for an original power of 
% $power_pp2*100$\%.

\section{Application}\label{sec:appli}

The \emph{Reproducibility Project: Psychology} \citep[\emph{RPP},][]{open2015},
the \emph{Experimental Economics Replication Project} \citep[\emph{EERP},][]{Camerer2016}, 
the \emph{Social Sciences Replication Project} \citep[\emph{SSRP},][]{Camerer2018} and the 
\emph{Experimental Philosophy Replicability Project} \citep[\emph{EPRP},][]{Cova2018}
are large-scale replication projects which aimed to replicate important 
discoveries published in journals from their respective fields.
Here we consider $138$ original studies 
considered to have a `positive' effect, \ie either significant or with a 'trend to significance' (\ie
non-significant with a $p$-value just slightly larger than the
significance threshold).
Several methods were used to assess replication success, such as the two-trials rule, 
meta-analysis of the original and replication effect estimates, or %%(equivalent 
%%to the weighted Stouffer's method), or 
compatibility of the replication effect estimates with a prediction interval based on the original results.

\begin{figure}[!h]
  \centering
\begin{knitrout}
\definecolor{shadecolor}{rgb}{0.969, 0.969, 0.969}\color{fgcolor}
\includegraphics[width=\maxwidth]{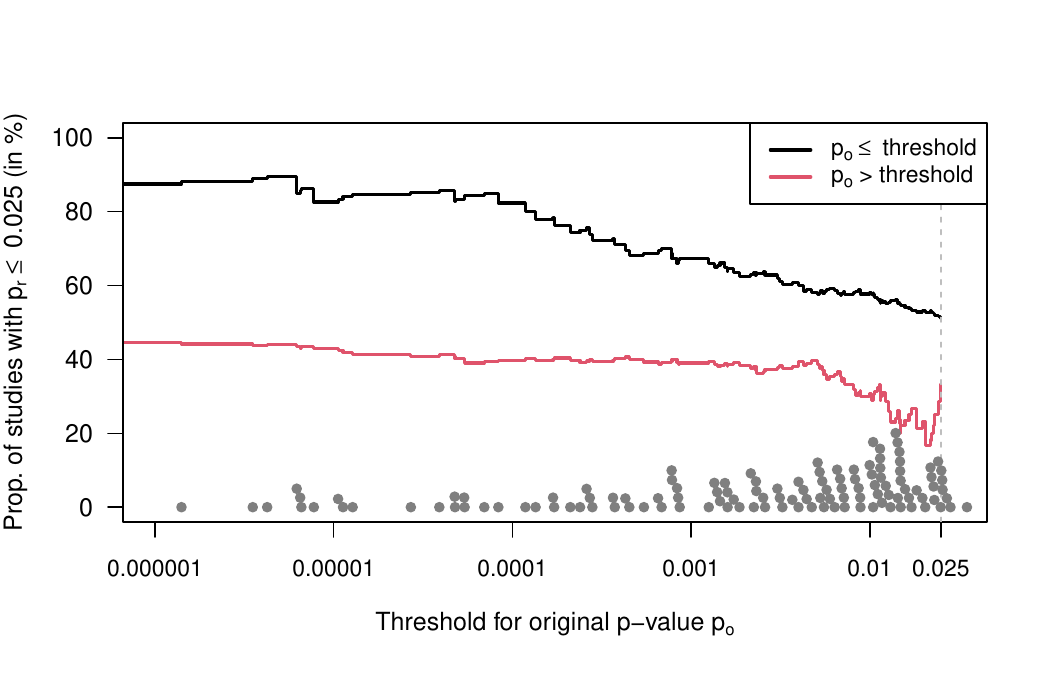} 
\end{knitrout}
\caption{Proportion of significant replication studies
  ($p_r \leq 0.025$) for original studies with $p$-value below and above
  the threshold, as a function of the threshold. The points at the bottom
  represent the original $p$-values $p_o$. There are
  17 study pairs with original $p$-value
  $p_o < 0.000001$ that are not shown.}
\label{fig:repRates}
\end{figure}

We grouped the 138 positive original studies into
significant ($p_o \leq$ threshold) and non-significant ($p_o >$ threshold) at
varying thresholds. The proportion of significant replication studies at the
one-sided 0.025 level ($p_r \leq \alpha = 0.025$) is then calculated for
each of the two groups and displayed in~Figure~\ref{fig:repRates}.
The proportion of significant replication studies is higher
for more convincing original studies (\ie, significant original studies at a smaller significance threshold).
This shows that more convincing original studies tend to replicate more than
less convincing original studies, and it therefore makes sense to be less
stringent with the former as does Edgington's method, but not the two-trials
  rule.

\begin{figure}[!h]
\centering
\begin{knitrout}
\definecolor{shadecolor}{rgb}{0.969, 0.969, 0.969}\color{fgcolor}
\includegraphics[width=\maxwidth]{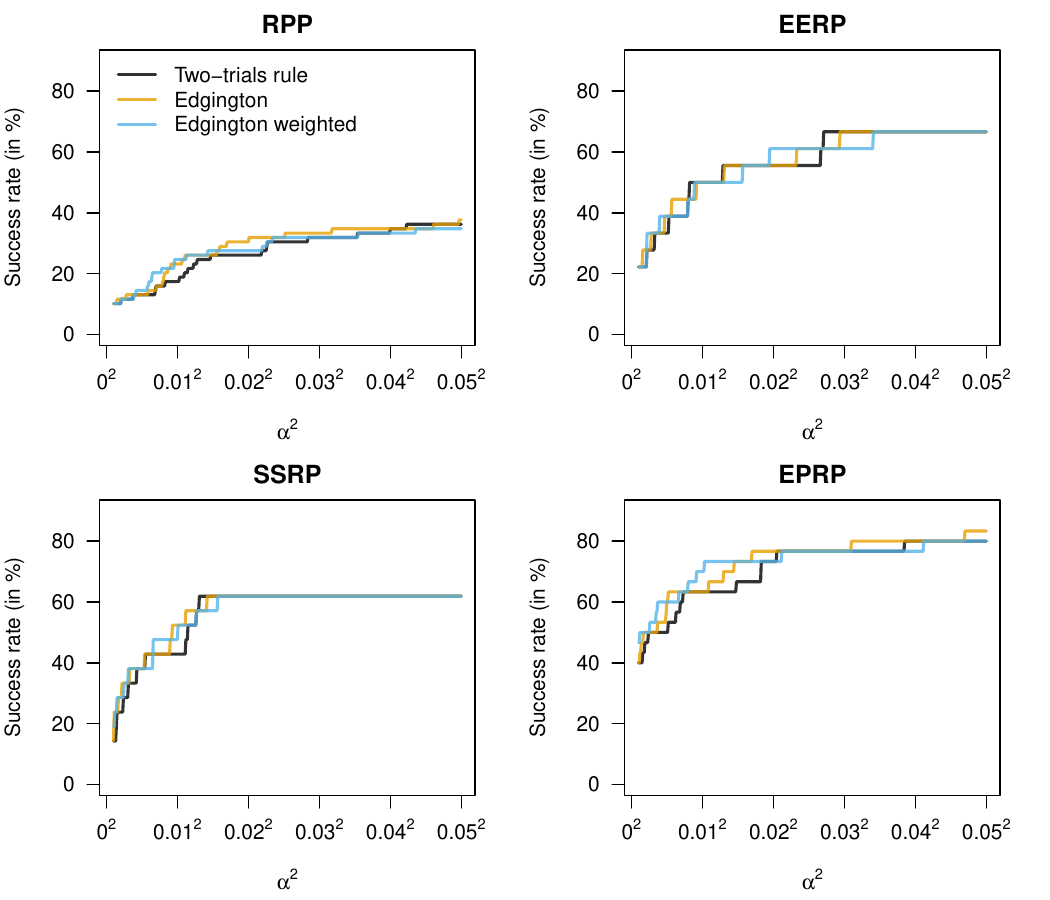} 
\end{knitrout}
\caption{Success rate of the \emph{RPP}, \emph{EERP}, \emph{SSRP} and 
\emph{EPRP} as a function of the overall \TOE rate $\alpha^2$ with the two-trials rule and
Edgington's method.}
\label{fig:appli}
\end{figure}

We then calculated the replication rates in the four projects with the
two-trials rule ($\pttr \leq \alpha^2$) and Edgington's method
($p_E \leq \alpha^2$ or $p_{E_w} \leq \alpha^2$) for varying levels
$\alpha^2$, see Figure~\ref{fig:appli}.  The replication rates
are similar with a tendency of larger success rates with Edgington's
method. This is in line with the larger project power of Edgington's
method discussed in Section \ref{sec:pp}.  For example, at level
  $\alpha^2 = 0.025^2$, the replication rate of the
two-trials rule and Edgington's method are 30.4\% vs 31.9\% in
the \emph{RPP}, 55.6\% vs
61.1\% in the \emph{EERP},
61.9\% for both in the \emph{SSRP} and
76.7\% for both in the \emph{EPRP}.  The
conclusions only differ for two study pairs: the original study by
\citet{Schmidt2008} and its replication in the \emph{RPP}, and the
original study by \citet{AmbrusGreiner2012} and its replication in the
\emph{EERP}.
As the original $p$-values 
$p_o = 0.028$ and
$p_o = 0.027$ are 
slightly larger than $\alpha = 0.025$ in both cases, the two-trials rule is not 
fulfilled. However, as $p_r < 0.0001$ and 
$p_r = 0.006$, 
respectively,
the sum $p_o + p_r$ does not exceed $b = 0.035$ and so replication 
success with Edgington's method is achieved for both study pairs.
Likewise, success is also achieved for the weighted method as $p_o + 2 \, p_r \leq 2 \, \alpha = 0.05$ in both cases.
%% \todo[inline]{SP: As an aside: For the Ambrus and Greiner study, conclusions would not change if we would use the slightly larger $p$-value from the table in the paper ($0.057/2 + 0.006 < 0.035$ and $0.057/2 + 2 \times 0.006 < 0.05)$}

\begin{figure}[!h]
\begin{knitrout}
\definecolor{shadecolor}{rgb}{0.969, 0.969, 0.969}\color{fgcolor}
\includegraphics[width=\maxwidth]{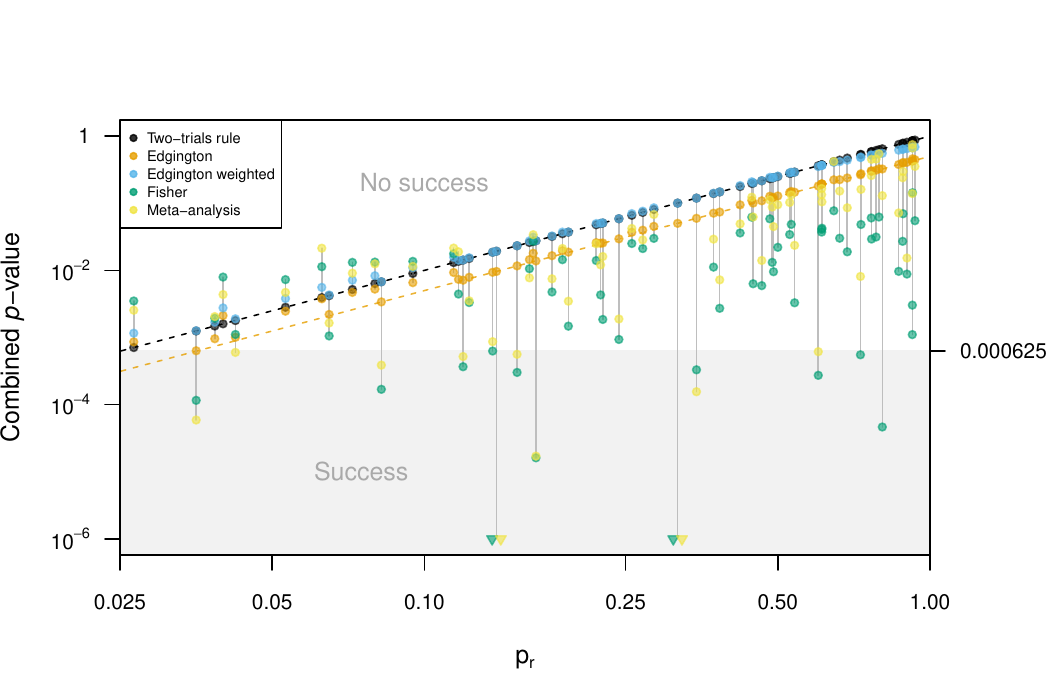} 
\end{knitrout}
\caption{Combined $p$-values $p_{\mbox{\tiny 2TR}}$, $p_E$, $p_{E_w}$, $p_F$,
  and $p_{\mbox{\tiny MA}}$ have been calculated from the original
  and replication $p$-values $p_o$ and $p_r$, respectively, for all replication studies
  considered in  the four replication projects.  They are plotted against the replication
  $p$-value $p_r$ for non-significant replication studies 
  ($p_r > 0.025$).
  Combined $p$-values in the grey area flag
  replication success at overall Type-I error rate $0.025^{2} = 0.000625$.  The
  dashed black line is the lower bound $p_r^2$ for $p_{\mbox{\tiny
      2TR}}$. % and $p_{E_w}$.
  The dashed orange line is the lower bound $p_r^2/2$ for
  $p_E \leq 1/2$. %% There is no such lower bound for $p_F$ and $p_{\mbox{\tiny  MA}}$.
  Fisher's ($p_F$) and meta-analytic ($p_{\mbox{\tiny
      MA}}$) $p$-values smaller than \ensuremath{10^{-6}} are marked with a
  triangle.  }
\label{fig:pCombined}
\end{figure}

We also
%% compared Fisher's method and the meta-analysis criterion in
%% the context of replication studies, so we 
computed the combined
$p$-values $p_{\mbox{\tiny 2TR}}$, $p_E$, $p_F$ and $p_{\mbox{\tiny
    MA}}$ with each of the four methods and plotted them against the
replication $p$-value $p_r$ for non-significant replication studies
(see Figure~\ref{fig:pCombined}). By construction, the combined
$p$-value from weighted Edgington's method is always larger than the
success threshold $\alpha^2=0.000625$ and this is also the case for the
unweighted version, where the smallest combined $p$-value
$p_E=0.000635$ is just slighty above
the threshold.  In contrast, with Fisher's method and the
meta-analysis criterion, replication success is often declared
although the replication $p$-value is quite large. There are even
three studies with $p_r > 0.5$ (so with an effect estimate in the
wrong direction) which achieve success at level $\alpha^2 =
0.000625$ with Fisher's method, and one
study with the meta-analysis criterion.  This illustrates that
Fisher's method and the meta-analysis criterion are not suited
as a replacement for the two-trials rule. 
%% in the context of replication studies.  %% However, the two-trials rule
%% and Edgington's method require both $p$-values to be smaller than
%% $\alpha = alpha$ and $b = round(b, 3)$, respectively,
%% and their combined $p$-values are always larger than $p_r^2$ and
%% $p_r^2/2$, respectively.  The combined $p$-value from weighted
%% Edgington's method is also always larger than $p_r^2$, except for
%% large values of $p_r$ where $E_w > 1$.

\section{Sample size calculation}\label{sec:design}

The sample size of the replication study is usually calculated based on conditional power, 
\ie such that
the power $1- \beta$ to reach a significant replication effect estimate 
reaches a certain value, usually 80\% or 90\%, assuming the original effect estimate is the true one. 
If, additionally, significance of the original study is required, 
this corresponds to the sample size calculation based on  the two-trials rule. 
In practice, a standard sample size calculation method is used where 
the minimal clinically important difference is replaced with the original 
effect estimate $\hat\theta_o$. For example, for a balanced two-sample $z$-test, 
the sample size per group in the 
replication study is calculated as
\begin{eqnarray}\label{eq:ss_abs}
n_r = \frac{2 \tau^2(z_{1 - \alpha} + z_{1 - \beta})^2}{\hat\theta_o^2} \, ,
\end{eqnarray}
where $\tau$ denotes the common standard deviation of the measurements, and $z_{1-u} = \Phi^{-1}(1-u)$ denotes the $1-u$ 
quantile of the standard normal distribution. We note that in some replication projects
\citep[\eg][]{Camerer2018} the 
original effect estimate $\hat\theta_o$ in \eqref{eq:ss_abs} is reduced by 25\% or even 50\%
to take into 
account its possible inflation \citep{ioannidis2008}.

It is also possible to express Equation~\eqref{eq:ss_abs} 
on the relative scale.  In that case, the required relative sample size 
$c = \sigma_o^2/\sigma_r^2 = n_r/n_o$ is calculated as
\begin{eqnarray}\label{eq:ss}
 c = \frac{\left(z_{1- \alpha} + z_{1- \beta}\right)^2}{z_o^2}   \, .
\end{eqnarray}

Sample size calculation based on \eqref{eq:ss_abs} or \eqref{eq:ss} is appropriate if
significance of the replication study at level $\alpha$ is the desired
criterion for replication success. If instead Edgington's method will
be used, the sample size calculation needs to be appropriately adapted
to ensure that the design of the replication study matches the analysis \citep{Anderson2022}.
To do so, the significance level $\alpha$ needs to be replaced with $b
- p_o$ in Equation~\eqref{eq:ss_abs} and~\eqref{eq:ss}, so now depends on the
$p$-value from the original study. 
A smaller sample size than with the two-trials rule is therefore required 
if $b - p_o > \alpha$, \ie
\begin{eqnarray}\label{eq:stringent}
p_o < b - \alpha = \sqrt{2} \, \alpha - \alpha = \alpha(\sqrt{2} - 1) \approx 0.01 \, .
\end{eqnarray}
The weighted version always requires a larger sample size than the two-trials rule, because the required significance level
is $\alpha - p_o/2 < \alpha$.

\begin{figure}[!h]
\begin{knitrout}
\definecolor{shadecolor}{rgb}{0.969, 0.969, 0.969}\color{fgcolor}
\includegraphics[width=\maxwidth]{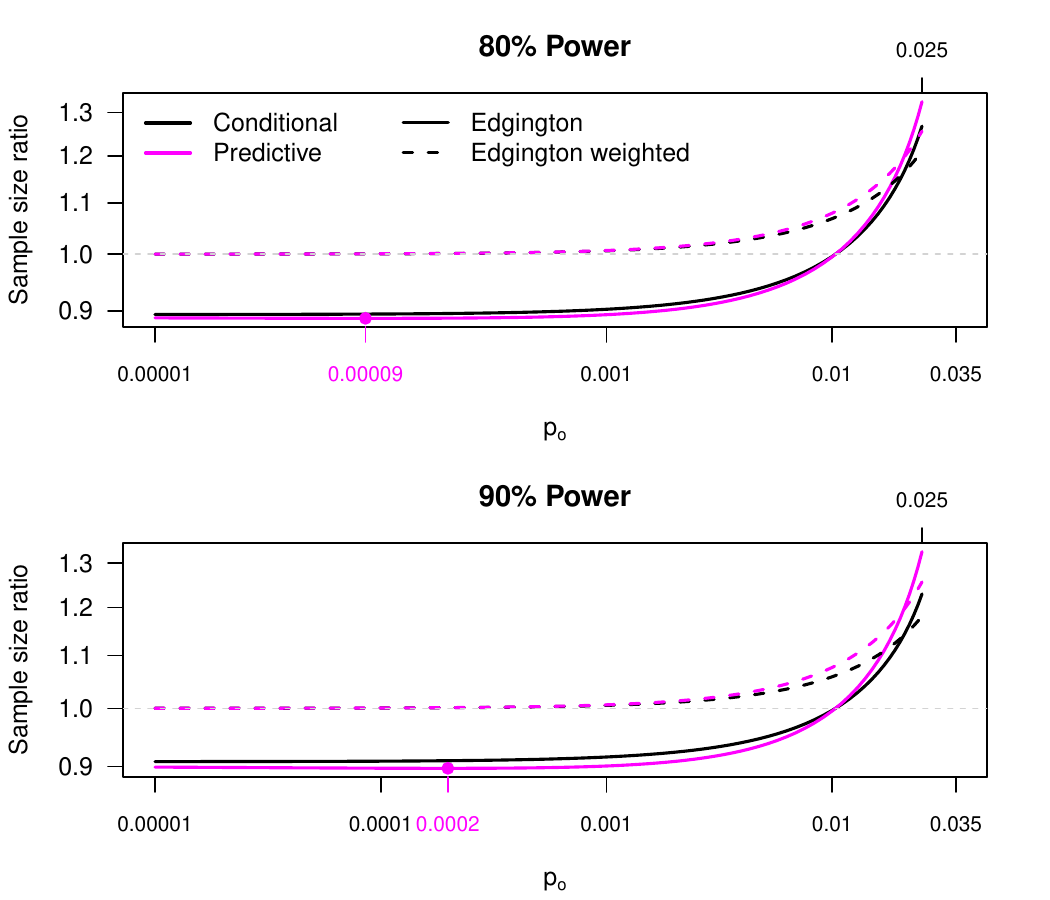} 
\end{knitrout}
\caption{Replication sample size ratio of Edgington's method compared
  to the two-trials rule to reach 80\% (top) and 90\% (bottom) power. For conditional power
  the sample size ratio is monotonically increasing, while it reaches
  a minimum at $p_o=0.00009$ for
  predictive power in the unweighted version. The corresponding sample
  size reduction is one minus the sample size ratio. The sample size ratio of the weighted
  version is always monotonically increasing and converges to 1 for $p_o \rightarrow 0$.}
\label{fig:ss_reduc}
\end{figure}

Figure~\ref{fig:ss_reduc} shows the sample size ratio
\[
\frac{\bigl(z_{1- b + p_o} + z_{1- \beta}\bigr)^2}{\bigl(z_{1- \alpha} + z_{1- \beta}\bigr)^2} \mbox{ resp.~}
\frac{\bigl(z_{1- \alpha + p_o/2} + z_{1- \beta}\bigr)^2}{\bigl(z_{1- \alpha} + z_{1- \beta}\bigr)^2} 
\]
of Edgington's method (unweighted and weighted) versus the two-trials rule for $\alpha=0.025$, a power of 80 and 90\% and
$p_o \in [0.00001, 0.025]$.
At  80\% power, the sample size calculated with unweighted Edgington's method can be up to 10.6\% smaller 
  than with the two-trials rule. At  90\% power, the sample size reduction looks very similiar with a maximum of 9.2\%.
%\todo[inline]{SP: would be interesting to actually see it because 80\% is also a very common target power}
  However, if $p_o > 0.01$, the required sample size with Edgington's 
  method is larger than with the two-trials rule. The weighted version always requires a larger sample size, but smaller than the unweighted version if $p_o$ is close to $\alpha=0.025$.

A drawback of conditional power is that it does not take the uncertainty of the 
original result into account and hence can lead to underpowered replication 
  studies.
One way to take into account uncertainty of the original result is to use 'predictive power'
instead \citep{MicheloudHeld2022}.
%
%% Figure~\ref{fig:ss} shows the required relative sample size $c$ to
%% reach a conditional or predictive power of $mypower*100$\%,
%% without shrinkage.
The relative sample size based on predictive power is generally larger
than based on conditional power.  The sample size reduction of
Edgington's method compared to the two-trials rule can be even more
pronounced and reaches a value of 11.2\% 
(10.3\%) at $p_o =
0.00009$
($p_o=0.0002$)
for 80\% (90\%) predictive rather than conditional power, see
Figure~\ref{fig:ss_reduc}.

%% \todo[inline]{@Leo, could you please add a few sentences about sequential application?}
%% \todo[inline, caption = {}]{Need to discuss sequential application of
%%   Edgington as described in \citet[Sec 4]{Held2023}.  Can be done in
%%   more generality than in \citet[Table 6]{Held2023} with $q \neq
%%   0.72$.  One could add a plot with $\alpha_2$, $\alpha_3$ as a
%%   function of $q$. One could investigate which value of $q$ is optimal
%%   wrt a certain criterion.}
%% \todo[inline]{Discuss bias through QRPs in the original study}

\section{Extensions to more than one replication study}\label{sec:moreThanOne}
It has been argued that a single replication study will often not be
sufficient and that more than one replication study is needed to
provide an unambiguous evaluation of replicability \citep{Hedges2019}.
Edgington's method can also be used if more than one replication study
is conducted. A simple approach would be to combine the replication
$p$-values into an overall replication $p$-value and then use
Edgington's method for one original and one replication $p$-value, as
introduced in this paper.  However, Edgington's method can also be
applied directly to the individual $p$-values, as we now illustrate in
the case of three studies (one original and two replications). Now the
sum $E_3=p_o + p_{r1} + p_{r2}$ of the three $p$-values needs to be
smaller than the new budget $b_3=0.16$ to ensure overall Type-I error
control at level $0.025^2$ \citep{Held2024}. An interesting aspect of
this approach is that it can be used to save resources if the
replication studies are conducted sequentially. Indeed, there will be
no point to conduct the second replication study if the sum of
$p$-values $E_2=p_o + p_{r1}$ from the original and the first
replication study is already larger than $b_3$.  Otherwise, a second
replication study at significance level $b_3 - E_2$ can be planned and
we would flag replication success if $E_3 = E_2 + p_{r2} \leq b_3$
holds. Such a sequential conduct of replication studies has been
suggested by \citet[p.~567]{Hedges2019} because ``a single initial
replication may be one effort in a sequence of replications, and as
researchers conduct additional subsequent replications, eventually a
preponderance of evidence will support more sensitive analyses.''

A refined version of this approach has been proposed in \citet[Section
  4]{Held2024}, allowing to stop for success already after the first
replication study. The approach is based on so-called alpha-spending
\citep{DeMetsLan1994}, distributing the overall Type-I error rate
$\alpha^2$ to the analysis after the first and after the second
replication study. 
Alpha-spending is a method originally proposed for interim analyses in
clinical trials.
%% Repeated significance testing in clinical trials is usually based on
%% group-sequential methods, where $p$-values are computed based on the
%% cumulative data available at each interim analysis.  However, there
%% are also proposals to calculate combined $p$-values based on the
%% available $p$-values from independent subsamples of the trial, as
%% defined by the timings of the interim analyses
%% \citep{MuellerSchaefer2001}.  
Specifically, Fisher's method has been
proposed for the evaluation of experiments with an adaptive interim
analysis %% , where $p_o$ and $p_r$ correspond to
based on the $p$-values %% for each
of two subsamples of the study %% and stopping boundaries have been calculated
\citep{Bauer1994}. 
Closer to our approach is the method by \citet{Chang2007}
who derives stopping rules
based on the sum of the $p$-values for each subsample of the trial.
%% , but apparently unaware of the original contribution by  \citet{Edgington1972}.

\begin{figure}
\begin{knitrout}
\definecolor{shadecolor}{rgb}{0.969, 0.969, 0.969}\color{fgcolor}
\includegraphics[width=\maxwidth]{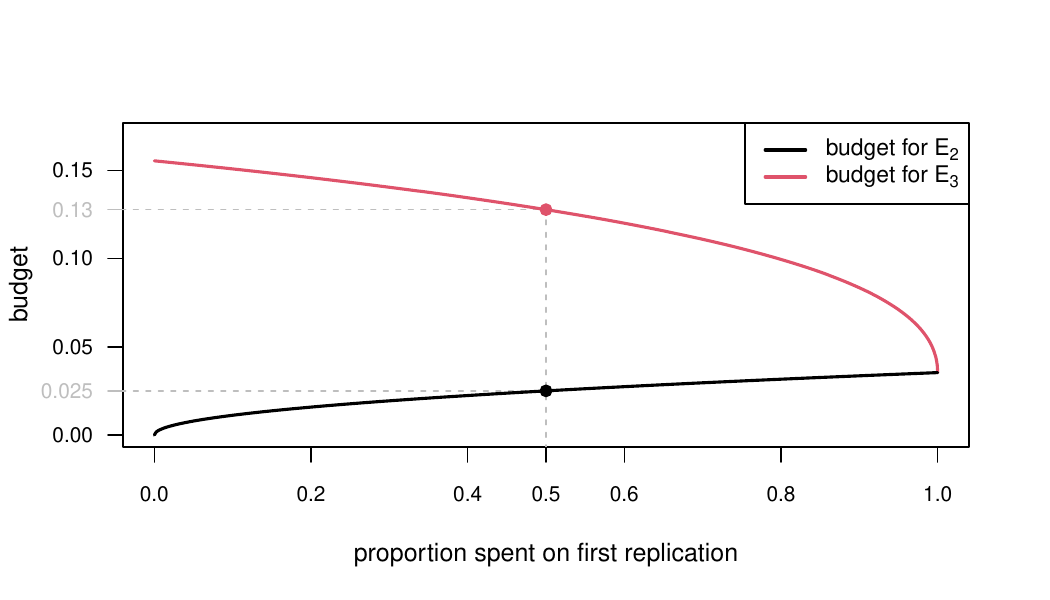} 
\end{knitrout}
\caption{Budget $b_2$ for $E_2=p_o + p_{r1}$ and $b_3$
  for $E_3=p_o + p_{r1} + p_{r2}$, respectively, depending on the proportion of
  $\alpha^2=0.025^2$ spent on the first replication study. The
  points denote the available budget if half of $\alpha^2$ is spent
  after the first replication.}
  \label{fig:alphaSpending}
\end{figure}

The application of the alpha-spending approach 
to the replication setting is illustrated in Figure
\ref{fig:alphaSpending}, which shows the budget $b_2$ for $E_2$ and
$b_3$ for $E_3$ depending on the proportion of $\alpha^2$ that is
spent on the analysis after the first replication. For example, if we
spend half of $\alpha^2$ on the first replication, we can stop for
replication success if $E_2=p_o + p_{r1} <
b_2=0.025$ holds. If this is not the case but at
least $E_2 < b_3=0.13$ holds, we would plan and
conduct a second replication study at significance level $b_3 - E_2$.
If eventually $E_3 = E_2 + p_{r2} \leq b_3=0.13$
holds, we have achieved success after the second replication
study. The combined procedure thus not only allows to stop for
replication success or failure after the first replication study, but
offers a third possibility to conduct a second replication study if
$b_2 = 0.025 < E_2 < b_3
=0.13$.  The approach could also be extended to
more than two replication studies and weights could also be
introduced. %% in order to downweight the original study.

\section{Discussion}\label{sec:discussion}

We propose to use the sum of the $p$-values, also known as Edgington's
method, instead of the two-trials rule in the assessment of
replication success. An unweighted and a weighted version are
considered.  In cases where it can be safely assumed that the original
study follows the same standards as the replication study
\citep[\textit{e.g.}][]{Protzko2023}, we recommend to use the
unweighted version. In cases where the original study may be
subject to questionable research practices or publication bias, we
recommend to give more weight to the replication study. The exact
choice of the weight depends on how much we distrust the original
study result. Both the unweighted and the weighted methods exactly
control the overall \TOE rate at level $\alpha^2$ and have an
acceptable bound on the conditional \TOE rate, namely $b =
0.035$ and $\alpha = 0.025$, respectively.
These numbers are for the conventional (but arbitrary) $\alpha =
0.025$ and a weight ratio of 2, and in principle other values could be
used.

The success bound for the replication $p$-value $p_r$ with Edgington's
method is not fixed at $\alpha = 0.025$ but varies between $0$
and $b = 0.035$ (unweighted) or between $0$ and $\alpha =
0.025$ (weighted), depending on the original $p$-value $p_o$.  Replication success is possible for
original studies that missed traditional statistical significance, as
long as $p_o \leq 0.035$ (unweighted) or $p_o \leq 2\,
\alpha = 0.05$ (weighted).  While these bounds are less
stringent than with the two-trials rule, they are also different with
the more elaborate sceptical $p$-value \citep{held2020}. The sceptical
$p$-value has been developed specifically for replication studies and
depends on the two $p$-values $p_o$ and $p_r$, but also on the variance
ratio, so treats original and replication study not as
exchangeable.  %% The golden recalibration \citep{held_etal2022b} has a
%% success bound of 0.062 for $\alpha=0.025$, while the recently
%% developed controlled version \citep{MicheloudHeld2023} does not impose
%% a restriction on the study-specific $p$-values, so is more
%% flexible than Edgington's method.
The controlled sceptical $p$-value \citep{MicheloudHeld2023} ensures
exact overall \TOE rate control, just as all methods discussed in this paper. It also
allows for replication success if the original study is non-significant and
can be used for sample size calculations. 
%% and its conditional \TOE rate now depends
%% on the relative sample size $c$. Sample size planning is is also
%% possible with this method, and then the conditional Type-I error rate
%% turns out to be always below 5\% for commonly used values (80 or 90\%) of
%% conditional or predictive power \citep[Section
%%   3.4]{MicheloudHeld2023}. 
However, the method is more complicated and therefore more difficult
to communicate. Edgington's method can be seen as a simple compromise between
the two-trials rule and the controlled sceptical $p$-value,
%% to the two-trials rule by
valuing the combined evidence from both studies while
ensuring that both studies support the alternative hypothesis.
%% While Edgington's
%% method seems like a promising replacement for the two-trials rule
Of course, researchers
may still want to quantify other aspects of replicability, such as statistical
consistency of original and replication effect estimates, for which other
methods, such as the $Q$-test, could be used \citep{Hedges2019}. In future work
we plan to conduct a simulation study to compare the operating characteristics
of Edgington's method with the sceptical $p$-value and alternative methods in
the presence of publication bias and other questionable research practices
\citep{Muradchanian2021,Freuli2023}.

One advantage of %% sample size calculation with
Edgington's method is
that it can be easily applied to non-normal or non-standard settings,
for example based on the $t$-test, a comparison of proportions, or the
log-rank test for survival data.  For example, for a sample size calculation based on the %% two-sample
$t$-test, the R function \texttt{power.t.test()} can be used to
calculate the required replication sample size $n_r$. The argument
\texttt{delta} needs to be set to the original effect estimate
$\hat\theta_o$ (perhaps incorporating some additional shrinkage) and
the argument \texttt{sig.level} to $\sqrt{2}\, \alpha - p_o$ (Edgington) or $\alpha - p_o/2$ (weighted Edgington) rather than
$\alpha$ (two-trials rule).

\section*{Data and software availability}

The R package \texttt{ReplicationSuccess} available on CRAN at
\url{https://CRAN.R-project.org/package=ReplicationSuccess}, has been
used for the sample size calculations.  The data of the \emph{RPP},
\emph{EERP}, \emph{SSRP} and \emph{EPRP} are available via the command
\texttt{data("RProjects")}. All $p$-values have been recalculated
based on Fisher's $z$-transformation as described in \citet[Supplementary Material]{PawelHeld2020}, see also
\texttt{help("RProjects")}. 
%\paragraph*{Code}
Code to reproduce the calculations in this manuscript is available at
\url{https://osf.io/uds2a/}. Functions to compute Edgington's combined $p$-value
(\texttt{pEdgington}) and associated power (\texttt{powerEdgington}) and sample
size calculations (\texttt{sampleSizeEdgington}) are available in the
development version of the \texttt{ReplicationSuccess} package which can be
installed with \texttt{remotes::install\_github(repo =
  "crsuzh/ReplicationSuccess", ref = "edgington")} (requires the
\texttt{remotes} package available on CRAN). We will make the new methods
available in the CRAN version soon.

%% \section*{Acknowledgments}
%% \todo[inline]{SP: Should this section be removed if there are no acknowledgments to make?}

\section*{Conflict of interest}
The authors have no conflict of interest to declare.

\section*{Appendix}
\appendix

% numbering of Figure 
\renewcommand{\thefigure}{A.\arabic{figure}}

\setcounter{figure}{0}

\section{Weighted sum of \textit{p}-values}\label{app:weightedSum}

The weighted sum of $p$-values \eqref{eq:Ew} with weights $w_o \leq
w_r$ can be written as
\[
E_w = q_o + q_r
\]
where $q_o \sim \Unif(0, w_o)$ and $q_r \sim \Unif(0, w_r)$ are independent uniform under the
intersection null hypothesis. The density function of $E_w$ can be
computed as the convolution of the densities of $q_o$ and $q_r$ and follows a trapezoidal distribution \citep{Dorp2003}:

\[
f(e_w) =  \left\{ \begin{array}{rl} \frac{e_w}{w_o w_r} &    \mbox{ if $0 < e_w \leq w_o$} \, , \\
    \frac{1}{w_r} & \mbox{ if $w_o < e_w \leq w_r$} \, ,  \\
    \frac{w_o + w_r - e_w}{w_o w_r} & \mbox{ if $w_r < e_w \leq w_o + w_r$} \, .
  \end{array} \right.
\]
The cumulative distribution function (cdf) of $E_w$ is therefore
\[
F(e_w) =  \left\{ \begin{array}{rl} \frac{e_w^2}{2\, w_o w_r} &    \mbox{ if $0 < e_w \leq w_o$} \, , \\
    \frac{1}{w_r} \left(e_w - \frac{w_o}{2} \right)& \mbox{ if $w_o < e_w \leq w_r$} \, ,  \\
    1 + \frac{1}{w_o w_r}\left(e_w(w_o + w_r) - \frac{(w_o + w_r)^{2}}{2} - \frac{e_w^{2}}{2} \right)
    & \mbox{ if $w_r < e_w \leq w_o + w_r$} \, .
  \end{array} \right.
\]
A valid combined $p$-value is obtained by plugging $E_w$ into the cdf: $p_{E_w} = F(E_w)$.

\section{Project power}\label{app:pp}
First assume that both effects are assumed the same, 
so $\theta_r = \theta_o$. As a result, $z_o \sim \Nor(\mu, 1)$
and $z_r \sim \Nor(\mu \sqrt{c}, 1)$.
The project power of the two-trials rule can be found in 
\citet[Section~3.3]{held_etal2022b}.
The project power of all other methods is calculated as
\[
\int \Pr(\mbox{Replication success} \given z_o) \phi(z_o - \mu) d z_o \, .
\]
Specifically, Edgington's method has project power
\begin{eqnarray}\label{eq:pp_E}
&& \int_{\Phi^{-1}(1 - b)}^{\infty}  
 \Pr(p_r \leq b - p_o) \phi(z_o - \mu) d z_o \, \nonumber \\
&& \int_{\Phi^{-1}(1 - b)}^{\infty}  \Pr(z_r \geq \Phi^{-1}(1-b + p_o) ) \phi(z_o - \mu) d z_o \, \nonumber \\
&& \int_{\Phi^{-1}(1 - b)}^{\infty}  \Pr(z_r \geq \Phi^{-1}(2-b -\Phi(z_o)) ) \phi(z_o - \mu) d z_o \, \nonumber \\
& = & \int_{\Phi^{-1}(1 - b)}^{\infty}
\left[1 - \Phi\left\{\Phi^{-1}\left(2 - b - \Phi(z_o)\right) - \mu \sqrt{c}\right\}\right] \phi(z_o - \mu) d z_o \, .
\end{eqnarray}

The project power of weighted Edgington's method with weight ratio
$\tilde w = w_r/w_o$ turns out to be
\begin{equation}\label{eq:pp_Ew}
\int_{\Phi^{-1}(1 - b_w/w_o)}^{\infty}
\left[1 - \Phi\left\{\Phi^{-1}\left(1 + 1/\tilde w - \sqrt{2/\tilde w}\, \alpha  - \Phi(z_o)/\tilde w\right) - \mu \sqrt{c}\right\}\right] \phi(z_o - \mu) d z_o \, .
\end{equation}

The project power of Fisher's method is 
\begin{eqnarray}\label{eq:pp_F}
\int_{0}^{\infty}
f(z_o) \, \phi(z_o - \mu) d z_o \mbox{ with } f(z_o) =  
\begin{cases}
1 - \Phi\left\{\Phi^{-1}\left(1 - \frac{c_F}{1 - \Phi(z_o)}\right)- \mu \sqrt{c}\right\} ,& \text{if } 1 - \Phi(z_o) \geq c_F\\
    1,              & \text{otherwise} \, ,
\end{cases}
\end{eqnarray}
and the project power of the meta-analysis criterion is 
\begin{eqnarray}\label{eq:pp_S}
\int_{0}^{\infty} \left[1 - \Phi\left\{\frac{\Phi^{-1}(1 - \alpha^2)\sqrt{c + 1} - z_o}{\sqrt{c}}\right\} - \mu \sqrt{c} \right ]
\phi(z_o - \mu) d z_o \, .
\end{eqnarray}
The project power of Edington's method converges to
\[
1 - \Phi\left(\Phi^{-1}(1-b) -  \Phi^{-1}(1-\alpha) - \Phi^{-1}(1-\beta)  \right)
\]
for $c \rightarrow \infty$, while the corresponding limit is
\[
1 - \Phi\left(\Phi^{-1}(1- b_w/w_o) -  \Phi^{-1}(1-\alpha) - \Phi^{-1}(1-\beta)  \right)
\]
%% \todo{LH: something wrong here?}
for weighted Edgington's method %% with weights $w_o=1$ and $w_r=2$ 
and 1 for the other two methods.

If $\theta_r = \theta_o/2$, the term $\mu\sqrt{c}$ in~\eqref{eq:pp_E}, \eqref{eq:pp_Ew}, \eqref{eq:pp_F}
and \eqref{eq:pp_S} needs to be divided by $2$.
\bibliographystyle{apalike}
\bibliography{antritt}
\end{document}